\documentclass[conference,10pt]{IEEEtran}

\usepackage{amsmath,amssymb,amsthm} 
\usepackage[dvips]{graphicx}
\usepackage{cite}
\usepackage{balance}

\newtheorem{definition}{Definition} 
\newtheorem{assumption}{Assumption} 
\newtheorem{proposition}{Proposition} 
\newtheorem{lemma}{Lemma} 
\newtheorem{theorem}{Theorem} 

\begin{document}

\sloppy

\title{A Potential Theory of General Spatially-Coupled Systems 
via a Continuum Approximation} 

\author{
  \IEEEauthorblockN{Keigo Takeuchi}
  \IEEEauthorblockA{
    Dept.\ Commun.\ Engineering \& Inf.\\ 
    University of Electro-Communications\\
    Tokyo 182-8585, Japan\\
    Email: ktakeuchi@uec.ac.jp
  }
  \and
  \IEEEauthorblockN{Toshiyuki Tanaka}
  \IEEEauthorblockA{
    Graduate School of Informatics \\ 
    Kyoto University \\ 
    Kyoto 606-8501, Japan \\
    Email: tt@i.kyoto-u.ac.jp
  }
  \and
  \IEEEauthorblockN{Kenta Kasai}
  \IEEEauthorblockA{
    Dept.\ Commun. \& Integrated Systems \\ 
    Tokyo Institute of Technology \\ 
    Tokyo 152-8550, Japan \\
    Email: kenta@comm.ss.titech.ac.jp
  }
}



\maketitle

\begin{abstract}
This paper analyzes general spatially-coupled (SC) systems with 
multi-dimensional coupling. A continuum approximation is used to derive 
potential functions that characterize the performance of the SC systems. 
For any dimension of coupling, it is shown that, if the boundary of the SC 
systems is fixed to the unique stable solution that minimizes the potential 
over all stationary solutions, the systems can approach the 
optimal performance as the number of coupled systems tends to infinity. 
\end{abstract}

\section{Introduction}
Kudekar et al.~\cite{Kudekar11} proved that spatial coupling can improve 
the belief-propagation (BP) performance of low-density parity-check (LDPC) 
codes up to the maximum-a-posteriori (MAP) performance. This phenomenon, 
called {\em threshold saturation}, has been observed in many other 
spatially-coupled (SC) systems, such as the MacKay-Neal and 
Hsu-Anastasopoulos codes~\cite{Kasai11}, 
code-division multiple-access (CDMA)~\cite{Takeuchi11,Takeuchi122,Schlegel11}, 
compressed sensing~\cite{Krzakala12,Donoho12}, and physical 
models~\cite{Hassani12}. Thus, threshold saturation via spatial coupling 
is believed to be a universal phenomenon. 

In order to prove the universality of threshold saturation, theoretical 
analyses have been performed for general SC systems with one-dimensional 
coupling~\cite{Takeuchi121,Yedla122,Kudekar12}. 
The methodologies are classified into those based on 
potential functions~\cite{Takeuchi121,Yedla122} and on extrinsic information 
transfer (EXIT) functions~\cite{Kudekar12}. 
Potential functions were also used for the analysis of threshold saturation 
in \cite{Donoho12,Hassani12}. The potential-based methodology 
has the advantage that the analysis of the BP performance is simplified.   
Yedla et al.~\cite{Yedla122} defined a potential function to specify the BP 
performance for general SC systems. However, they presented no derivation of 
the potential function. One purpose of this paper is 
to present a systematic derivation of the potential function. 

The derivation is based on the continuum approximation used in  
\cite{Takeuchi121,Takeuchi122}. The continuum approximation can be naturally 
extended to the case of multi-dimensional coupling: The previous analysis 
for SC scalar systems with one-dimensional coupling is generalized to the 
case of SC vector systems with multi-dimensional coupling. 
Hereafter, general SC vector systems with multi-dimensional coupling is 
simply referred to as generalized SC (GSC) systems. Multi-dimensional 
coupling can provide robustness of convergence against burst errors. 
See \cite{Ohashi13} for the details.  

The main contributions of this paper are summarized as follows: 
(i) The potential function defined in \cite{Yedla122} is systematically 
derived via the continuum approximation. 
(ii) Multi-dimensional coupling is shown to provide the same improvement of 
the BP performance as one-dimensional coupling. 

\section{System Model} 
\subsection{Notation} 
For integers $i$ and $j$ ($>i$), $[i:j]$ denotes the set 
$\{i,i+1,\ldots,j\}$ of lattice points. 
The symbols~$\delta_{ab}$, $\delta_{a}^{b}$, $\delta^{ab}$ 
represent the Kronecker delta. 
As defined below, the state of a GSC system with $K$-dimensional coupling is 
represented by two $N$-dimensional vector fields 
$\boldsymbol{u}(\boldsymbol{x},t) 
= \{u^{a}(\boldsymbol{x},t): a=1,\ldots,N\}$ and 
$\boldsymbol{v}(\boldsymbol{x},t) = \{v_{a}(\boldsymbol{x},t): a=1,\ldots,N\}$ 
on $\mathbb{R}^{K}\times[0,\infty)$. Roman alphabet is used for the indices 
of the elements of the vector fields, whereas Greek alphabet is for the 
spatial vector  $\boldsymbol{x}=\{x^{\alpha}:\alpha=1,\ldots,K\}$. 
The differential 
operators $\partial/\partial u^{a}$ and $\partial/\partial v_{b}$ are 
abbreviated to $\partial_{a}$ and $\partial^{b}$, respectively.  
The gradients $\{\partial_{a}:a=1,\ldots,N\}$ and 
$\{\partial^{a}:a=1,\ldots,N\}$ are denoted by $\nabla$ and $\tilde{\nabla}$, 
respectively. One the other hand, $\partial/\partial\boldsymbol{x}$ represents 
the gradient $\{\partial/\partial x^{\alpha}:\alpha=1,\ldots,K\}$ 
for the spatial variables. Furthermore, the Einstein summation convention is 
used: When an index appears twice in a single term, the summation is taken 
over all values of the index. For example, $u^{a}v_{a}=\sum_{a=1}^{N}u^{a}v_{a}$. 

\subsection{Uncoupled System} 
For two scalar fields $F$ and $G$ on $\mathbb{R}^{N}$, 
let $\mathcal{D}\subset\mathbb{R}^{N}$ and 
$\tilde{\mathcal{D}}\subset\mathbb{R}^{N}$ denote the images of the gradients 
$\tilde{\nabla} F$ and $\nabla G$, respectively. The dimension $N$ corresponds 
to the number of parameters required for describing asymptotic performance 
of the BP algorithm for a system with no coupling. We assume that asymptotic 
performance is characterized by the coupled density-evolution (DE) equations 
with respect to $\boldsymbol{u}(t)=\{u^{a}(t):a=1,\ldots,N\}$ and 
$\boldsymbol{v}(t)=\{v_{a}(t):a=1,\ldots,N\}$, 
\begin{equation} \label{DE1} 
\boldsymbol{u}(t+1) 
= \tilde{\nabla} F(\boldsymbol{v}(t)) \in\mathcal{D}, 
\end{equation}
\begin{equation} \label{DE2} 
\boldsymbol{v}(t) 
= \nabla G(\boldsymbol{u}(t))\in\tilde{\mathcal{D}}, 
\end{equation}
where the time index~$t$ corresponds to the number of iterations for the 
BP algorithm. The asymptotic performance in iteration~$t$ is assumed to be 
characterized by a deterministic scalar function $P(\cdot)$ of the state 
$\boldsymbol{u}(t)$ that starts from an appropriate initial state. 

\begin{assumption} \label{assumption1} 
The two scalar fields $F$ and $G$ are thrice continuously differentiable on 
$\tilde{\mathcal{D}}$ and $\mathcal{D}$, respectively.  
Let $\mathcal{S}\subset\mathcal{D}$ and $\tilde{\mathcal{S}}\subset
\tilde{\mathcal{D}}$ denote the sets of first and second elements of 
all fixed-points (FPs) 
$(\boldsymbol{u},\boldsymbol{v})$ for the DE equations~(\ref{DE1}) 
and (\ref{DE2}), respectively. The Hesse matrix $f^{ab}(\boldsymbol{v})
=\partial^{a}\partial^{b}F(\boldsymbol{v})$ is non-singular for all 
$\boldsymbol{v}\in\tilde{\mathcal{D}}_{0}
=\tilde{\mathcal{D}}\backslash\tilde{\mathcal{S}}$, and the quadratic form 
$y_{a}y_{b}f^{ab}(\boldsymbol{v})$ is bounded below for all 
$\{y_{a}\}\in\tilde{\mathcal{D}}$ and $\boldsymbol{v}\in\tilde{\mathcal{D}}$.  
On the other hand, the Hesse matrix 
$g_{ab}(\boldsymbol{u})=\partial_{a}\partial_{b}G(\boldsymbol{u})$ are 
positive (resp.\ non-negative) definite for 
$\boldsymbol{u}\in\mathcal{D}_{0}
=\mathcal{D}\backslash\mathcal{S}$ 
(resp.\ $\boldsymbol{u}\in\mathcal{D}$). 
\end{assumption}

Examples that satisfy Assumption~\ref{assumption1} are 
MN and HA codes~\cite{Kasai11} as well as LDPC 
codes~\cite{Kudekar11} and CDMA~\cite{Takeuchi11,Takeuchi122,Schlegel11}. 
Note that Assumption~\ref{assumption1} is different from that 
in \cite{Yedla122}: The positive-definiteness of $g_{ab}(\boldsymbol{u})$ is 
assumed in this paper, whereas the positivity of 
its elements is postulated in \cite{Yedla122}. 

In order to investigate the convergence property of the state 
$\boldsymbol{u}(t)$, we define two potential functions, one of which was 
originally defined in \cite{Yedla122}. They are equivalent 
to the so-called trial entropy for LDPC codes~\cite{Yedla122} and to the 
free energy for CDMA~\cite{Takeuchi122}, characterizing the MAP performance. 


\begin{definition}[Potential] 
Let $D(\boldsymbol{u},\boldsymbol{v})$ denote the divergence   
\begin{equation} \label{divergence} 
D(\boldsymbol{u},\boldsymbol{v}) 
= G(\boldsymbol{u}) + F(\boldsymbol{v}) - u^{a}v_{a}. 
\end{equation}
The potential function $V(\boldsymbol{u})$ is defined as  
\begin{equation} \label{potential} 
V(\boldsymbol{u}) = - D(\boldsymbol{u},\nabla G(\boldsymbol{u})).  
\end{equation} 
On the other hand, the dual potential function $\tilde{V}(\boldsymbol{v})$ 
is defined by 
\begin{equation} \label{dual_potential} 
\tilde{V}(\boldsymbol{v}) 
= - D(\tilde{\nabla} F(\boldsymbol{v}),\boldsymbol{v}). 
\end{equation}
\end{definition}
\begin{assumption} \label{assumption2} 
The potential functions~(\ref{potential}) and (\ref{dual_potential}) 
are bounded below. 
\end{assumption}
Assumption~\ref{assumption2} is a sufficient condition for guaranteeing 
the convergence of the state $\boldsymbol{u}(t)$ toward a FP 
$\boldsymbol{u}^{*}$ as $t\to\infty$. The following proposition implies that 
the two potential functions have the same information about FPs.  

\begin{proposition} \label{proposition1} 
The $2$-tuple $(\boldsymbol{u}^{*},\boldsymbol{v}^{*})$ is a FP 
of the DE equations~(\ref{DE1}) and (\ref{DE2}) if and only if 
$\boldsymbol{u}^{*}$ (resp.\ $\boldsymbol{v}^{*}$) is a stationary solution of 
the potential~(\ref{potential}) (resp.\ (\ref{dual_potential})). 
Furthermore, $V(\boldsymbol{u}^{*})=\tilde{V}(\boldsymbol{v}^{*})$ 
at any FP $(\boldsymbol{u}^{*},\boldsymbol{v}^{*})$. 
\end{proposition}
\begin{IEEEproof}
Calculating the gradient of the potential~(\ref{potential}) 
with (\ref{divergence}) yields 
\begin{equation} \label{gradient} 
\partial_{a}V(\boldsymbol{u}) 
= g_{ab}(\boldsymbol{u})\{u^{b} - \partial^{b}F(\nabla G(\boldsymbol{u}))\}. 
\end{equation}
Since the Hesse matrix $\{g_{ab}(\boldsymbol{u})\}$ of $G$ is non-singular 
except for the FPs $\boldsymbol{u}^{*}\in\mathcal{S}$, 
$\partial_{a}V(\boldsymbol{u}^{*})=0$ is equivalent to 
$\boldsymbol{u}^{*} = \tilde{\nabla} F(\boldsymbol{v}^{*})$ with 
$\boldsymbol{v}^{*}=\nabla G(\boldsymbol{u}^{*})$. Similarly, we find that 
$\partial^{a}\tilde{V}(\boldsymbol{v}^{*})=0$ is equivalent to the condition 
that $(\boldsymbol{u}^{*},\boldsymbol{v}^{*})$ is a FP.  
Furthermore, we have 
\begin{equation}
\tilde{V}(\nabla G(\boldsymbol{u}^{*})) 
= - D(\tilde{\nabla} F(\nabla G(\boldsymbol{u}^{*})),
\nabla G(\boldsymbol{u}^{*}))
= V(\boldsymbol{u}^{*}), 
\end{equation} 
where $\boldsymbol{u}^{*}=\tilde{\nabla} F(\nabla G(\boldsymbol{u}^{*}))$ has 
been used. This implies $\tilde{V}(\boldsymbol{v}^{*})
=V(\boldsymbol{u}^{*})$. 
\end{IEEEproof} 

Let us investigate the convergence property of the state. 
We use the approximation 
$\boldsymbol{u}(t+1)-\boldsymbol{u}(t)\approx d\boldsymbol{u}/dt$ 
for (\ref{DE1}) to obtain 
\begin{equation} 
\frac{du^{a}}{dt} 
\approx - \{u^{a} - \partial^{a}F(\nabla G(\boldsymbol{u}))\} 
= - g^{ab}(\boldsymbol{u})\partial_{b}V(\boldsymbol{u}), 
\label{continuous_system} 
\end{equation}
for $\boldsymbol{u}\in\mathcal{D}_{0}$, where 
we have used (\ref{gradient}). In (\ref{continuous_system}), 
$\{g^{ab}(\boldsymbol{u})\}$ denotes the inverse matrix of the Hesse 
matrix $\{g_{ab}(\boldsymbol{u})\}$. 
It is straightforward to confirm that the potential~(\ref{potential}) is 
a Lyapunov function for the continuous-time system~(\ref{continuous_system}). 
\begin{equation}
\frac{dV}{dt}(\boldsymbol{u}(t)) 
= \partial_{a}V\frac{du^{a}}{dt} 
= - \partial_{a}V(\boldsymbol{u}(t))\partial_{b}V(\boldsymbol{u}(t))
g^{ab}(\boldsymbol{u}(t)),  
\end{equation}
which is negative for all $\boldsymbol{u}(t)\in\mathcal{D}_{0}$,  
and zero only when $\boldsymbol{u}(t)$ is at a stationary solution of the 
potential~(\ref{potential}). Since the potential~(\ref{potential}) is 
bounded below, this observation implies that the state 
$\boldsymbol{u}(t)$ for the continuous-time system~(\ref{continuous_system}) 
converges to a stationary solution of the potential~(\ref{potential}) 
as $t\to\infty$. 

\begin{figure}[t]
\begin{center}
\includegraphics[width=\hsize]{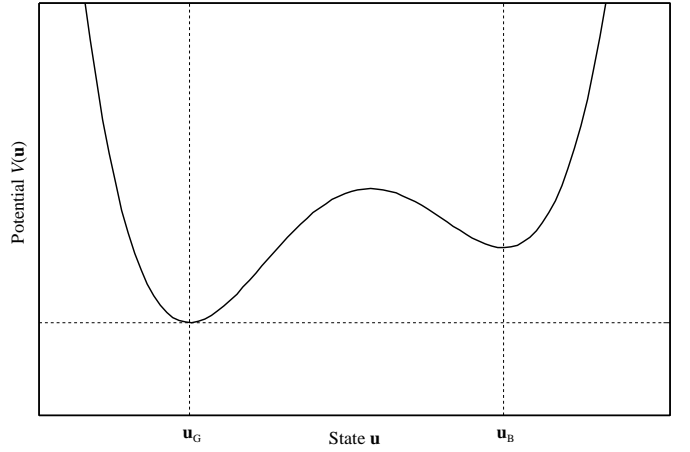}
\end{center}
\caption{
Shape of the potential postulated in this paper. 
}
\label{fig1} 
\end{figure}

\subsection{General Spatially-Coupled System}
Let us consider the case of double-well potential shown in Fig.~\ref{fig1}. 
The potential $V(\boldsymbol{u}_{\mathrm{G}})$ at the left stable solution 
$\boldsymbol{u}_{\mathrm{G}}$ is lower than that at the right 
stable solution $\boldsymbol{u}_{\mathrm{B}}$. Furthermore, the performance 
$P(\boldsymbol{u}_{\mathrm{G}})$ is assumed to be better than  
$P(\boldsymbol{u}_{\mathrm{B}})$. Note that there is 
no relationship between $P(\boldsymbol{u})$ and $V(\boldsymbol{u})$. 
We hereafter refer to $\boldsymbol{u}_{\mathrm{G}}$ and 
$\boldsymbol{u}_{\mathrm{B}}$ as the good and bad solutions, respectively.  
The state $\boldsymbol{u}(t)$ for the BP algorithm should converge to 
the bad stable solution $\boldsymbol{u}_{\mathrm{B}}$, since the BP algorithm 
starts from a bad initial state. On the other hand, the MAP algorithm 
may achieve the better performance $P(\boldsymbol{u}_{\mathrm{G}})$, 
which implies the suboptimality of the BP algorithm. Spatial coupling is 
a method for helping the state climb the potential barrier and arrive at 
the good stable solution $\boldsymbol{u}_{\mathrm{G}}$.  

Suppose that a GSC system with $K$-dimensional coupling of size~$L$ and 
width~$W$ is characterized by the DE equations, 
\begin{equation} \label{DE1_GSC}
\boldsymbol{u}\left(
 \frac{\boldsymbol{l}}{L},t+1
\right) 
= \left\langle
 \tilde{\nabla} F\left(
  \boldsymbol{v}\left(
   \frac{\boldsymbol{l}+\boldsymbol{m}}{L},t
  \right)
 \right)
\right\rangle_{\boldsymbol{m}}, 
\end{equation}
\begin{equation} \label{DE2_GSC} 
\boldsymbol{v}\left(
 \frac{\boldsymbol{l}}{L},t
\right) 
= \left\langle
 \nabla G\left(
  \boldsymbol{u}\left(
   \frac{\boldsymbol{l}-\boldsymbol{m}}{L},t
  \right)
 \right)
\right\rangle_{\boldsymbol{m}},  
\end{equation}
for all spatial positions~$\boldsymbol{l}\in[-L+1:L-1]^{K}$, with 
\begin{equation}
\langle f(\boldsymbol{m}) \rangle_{\boldsymbol{m}} 
= \frac{1}{(2W+1)^{K}}\sum_{\boldsymbol{m}\in[-W:W]^{K}}f(\boldsymbol{m}).  
\end{equation}
We impose the boundary condition $\boldsymbol{u}(\boldsymbol{l},t)
=\boldsymbol{u}_{\mathrm{G}}$ for all 
$\boldsymbol{l}\in\mathbb{Z}^{K}\backslash [-L+1:L-1]^{K}$. 
This condition corresponds to informing the detector about the true solutions 
for the systems at the boundary in advance. 

The idea of spatial coupling is explained as follows: The correct information 
at the boundary may spread over the whole system via coupling, 
regardless of size~$L$. Since the influence of the boundary is negligible 
as $L\to\infty$, the loss due to informing the detector about the true 
solutions is also negligible.

\section{Continuum Approximation}
In order to investigate the convergence property of the DE 
equations~(\ref{DE1_GSC}) and (\ref{DE2_GSC}), we use the 
continuum approximation with respect to the spatial 
variable $\boldsymbol{x}=\boldsymbol{l}/L$ as $L\to\infty$. 
For notational convenience, the argument $(\boldsymbol{x},t)$ 
of functions is omitted. 
\begin{lemma}
Let 
\begin{equation}
M = \frac{1}{L^{2}(2W+1)}\sum_{m=-W}^{W}m^{2}. 
\end{equation}
As $L\to\infty$, the DE equations~(\ref{DE1}) and (\ref{DE2}) are respectively 
approximated by the spatially continuous equations  
\begin{IEEEeqnarray}{r} 
u^{a}(\boldsymbol{x},t+1) 
= \partial^{a}F(\boldsymbol{v})  
+ \frac{M}{2}\sum_{\alpha=1}^{K}\left(
 f^{ab}(\boldsymbol{v})\frac{\partial^{2}v_{b}}{\partial {x^{\alpha}}^{2}}
\right. \nonumber \\ 
\left.
 + \partial^{a}\partial^{b}\partial^{c}F(\boldsymbol{v}) 
 \frac{\partial v_{b}}{\partial x^{\alpha}}
 \frac{\partial v_{c}}{\partial x^{\alpha}}
\right) + o(L^{2}), \label{continuous_DE1} 
\end{IEEEeqnarray} 
\begin{IEEEeqnarray}{r} 
v_{a} = \partial_{a}G(\boldsymbol{u}) 
+ \frac{M}{2}\sum_{\alpha=1}^{K}\left(
 g_{ab}(\boldsymbol{u})\frac{\partial^{2}u^{b}}{\partial {x^{\alpha}}^{2}}
\right. \nonumber \\ 
\left.
 + \partial_{a}\partial_{b}\partial_{c}G(\boldsymbol{u}) 
 \frac{\partial u^{b}}{\partial x^{\alpha}}
 \frac{\partial u^{c}}{\partial x^{\alpha}}
\right) + o(L^{2}). \label{continuous_DE2}
\end{IEEEeqnarray} 
\end{lemma} 
\begin{IEEEproof}
The proof is based on a straightforward Taylor expansion with respect to 
$1/L$, and therefore omitted.  
\end{IEEEproof}

Note that cross terms with respect to the spatial variables vanish because 
of $\langle \boldsymbol{m} \rangle_{\boldsymbol{m}}=\boldsymbol{0}$. 
In terms of differential geometry, the variable $\boldsymbol{u}$ may be 
regarded as the coordinates associated with a local coordinate system in 
a differential manifold with an affine connection. The coefficient 
$\partial_{a}\partial_{b}\partial_{c}G(\boldsymbol{u})$ 
in (\ref{continuous_DE2}) corresponds to the coefficient of the 
affine connection for the local coordinate system. 
The same interpretation holds for $\boldsymbol{v}$ and 
$\partial^{a}\partial^{b}\partial^{c}F(\boldsymbol{v})$.  
A direct calculation implies that the two manifolds for 
$\boldsymbol{u}$ and $\boldsymbol{v}$ are flat, i.e.\ 
the torsion and curvature tensors are everywhere zero. Consequently, 
we can use affine coordinate systems suitable for representing flat 
manifolds. It is known that the coefficients of the 
affine connections vanish everywhere for affine coordinate systems. 

\begin{theorem} \label{theorem1} 
Consider the affine coordinates 
$(\tilde{\boldsymbol{u}},\tilde{\boldsymbol{v}})$ that is defined by 
the twice continuously differentiable mapping from 
$(\boldsymbol{u},\boldsymbol{v})\in
\mathcal{D}_{0}\times\tilde{\mathcal{D}}_{0}$ 
onto $(\tilde{\boldsymbol{u}},\tilde{\boldsymbol{v}})
\in\mathcal{D}_{0}\times\tilde{\mathcal{D}}_{0}$,   
\begin{equation} \label{transform} 
(\tilde{\boldsymbol{u}},\tilde{\boldsymbol{v}}) 
= (\tilde{\nabla} F(\boldsymbol{v}),\nabla G(\boldsymbol{u})).  
\end{equation}
Fix $(\boldsymbol{x},t)$ and suppose that $(\boldsymbol{u},\boldsymbol{v})
\in\mathcal{D}_{0}\times\tilde{\mathcal{D}}_{0}$ in a neighborhood of 
$(\boldsymbol{x},t)$. Then, 
the equations~(\ref{continuous_DE1}) and (\ref{continuous_DE2}) reduce to the 
two decoupled equations with respect to the affine coordinates 
$\tilde{\boldsymbol{v}}=\{\tilde{v}_{a}:a=1,\ldots,N\}$ and 
$\tilde{\boldsymbol{u}}=\{\tilde{u}^{a}:a=1,\ldots,N\}$,  
\begin{IEEEeqnarray}{rl} 
&\Phi^{a}(\tilde{\boldsymbol{v}}(\boldsymbol{x},t+1)) 
- \Phi^{a}(\tilde{\boldsymbol{v}}(\boldsymbol{x},t)) 
\nonumber \\ 
=& - \frac{\partial}{\partial\tilde{v}_{a}}
V(\boldsymbol{\Phi}(\tilde{\boldsymbol{v}})) 
+ M\mathfrak{C}^{a}(\tilde{\boldsymbol{v}}) 
+ o(L^{2}), \;\;\; \label{PDE} 
\end{IEEEeqnarray}
\begin{IEEEeqnarray}{rl} 
&\Psi_{a}(\tilde{\boldsymbol{u}}(\boldsymbol{x},t+1)) 
- \Psi_{a}(\tilde{\boldsymbol{u}}(\boldsymbol{x},t)) 
\nonumber \\ 
=& - \frac{\partial}{\partial\tilde{u}^{a}}
\tilde{V}(\boldsymbol{\Psi}(\tilde{\boldsymbol{u}})) 
+ M\tilde{\mathfrak{C}}_{a}(\tilde{\boldsymbol{u}}) 
+ o(L^{2}), \;\;\; \label{dual_PDE} 
\end{IEEEeqnarray}
with 
\begin{equation} \label{coupling_operator} 
\mathfrak{C}^{a}(\tilde{\boldsymbol{v}})  
= \sum_{\alpha=1}^{K}\left(
 f^{ab}(\tilde{\boldsymbol{v}})
 \frac{\partial^{2}\tilde{v}_{b}}{\partial {x^{\alpha}}^{2}} 
 + \frac{1}{2}\frac{\partial f^{ab}}{\partial v_{c}}(\tilde{\boldsymbol{v}}) 
 \frac{\partial\tilde{v}_{b}}{\partial x^{\alpha}}  
 \frac{\partial\tilde{v}_{c}}{\partial x^{\alpha}}
\right), 
\end{equation} 
\begin{equation} \label{dual_coupling_operator} 
\tilde{\mathfrak{C}}_{a}(\tilde{\boldsymbol{u}})  
= \sum_{\alpha=1}^{K}\left(
 g_{ab}(\tilde{\boldsymbol{u}})
 \frac{\partial^{2}\tilde{u}^{b}}{\partial {x^{\alpha}}^{2}} 
 + \frac{1}{2}\frac{\partial g_{ab}}{\partial u_{c}}(\tilde{\boldsymbol{u}}) 
 \frac{\partial\tilde{u}^{b}}{\partial x^{\alpha}}  
 \frac{\partial\tilde{u}^{c}}{\partial x^{\alpha}}
\right). 
\end{equation} 
In the first terms on the right-hand sides (RHSs) of (\ref{PDE}) and 
(\ref{dual_PDE}), 
the potential functions are given by (\ref{potential}) and 
(\ref{dual_potential}), respectively. Furthermore,  
$\boldsymbol{\Phi}(\tilde{\boldsymbol{v}}) 
=\{\Phi^{a}(\tilde{\boldsymbol{v}}):a=1,\ldots,N\}$ and 
$\boldsymbol{\Psi}(\tilde{\boldsymbol{u}}) 
=\{\Psi_{a}(\tilde{\boldsymbol{u}}):a=1,\ldots,N\}$ denote the 
inverse mapping of $\tilde{\boldsymbol{v}}=\nabla G(\boldsymbol{u})$ 
and $\tilde{\boldsymbol{u}}=\tilde{\nabla}F(\boldsymbol{v})$, respectively. 
\end{theorem}

The first terms on the RHSs of (\ref{PDE}) and (\ref{dual_PDE}) are potential 
terms determined by the properties of the corresponding uncoupled system, 
whereas the second terms represent the effect of spatial coupling. 

\begin{IEEEproof}
The change of coordinates~(\ref{transform}) transforms the second term 
on the RHS of (\ref{continuous_DE2}) into 
\begin{equation} \label{continuous_DE2_tmp} 
v_{a} 
= \partial_{a}G(\boldsymbol{u})
+ \frac{M}{2}\sum_{\alpha=1}^{K}\left(
 \frac{\partial^{2}\tilde{v}_{a}}{\partial {x^{\alpha}}^{2}}
 +  \Gamma_{a}^{bc}\frac{\partial\tilde{v}_{b}}{\partial x^{\alpha}}
 \frac{\partial \tilde{v}_{c}}{\partial x^{\alpha}}
\right) + o(L^{2}), 
\end{equation} 
with
\begin{equation}
\Gamma_{a}^{bc} = 
g^{db}(\boldsymbol{u})g^{ec}(\boldsymbol{u}) 
\partial_{a}\partial_{d}\partial_{e}G(\boldsymbol{u}) 
+ g_{ad}(\boldsymbol{u})\frac{\partial}{\partial\tilde{v}_{c}}
g^{db}(\boldsymbol{u}). 
\end{equation}
Using a formula obtained by differentiating the identity  
\begin{equation} \label{inverse_Hesse} 
g_{ad}(\boldsymbol{u})g^{db}(\boldsymbol{u})
= \delta_{a}^{b},  
\end{equation}
with respect to $\tilde{v}_{c}$, we find 
\begin{equation} \label{connection} 
\Gamma_{a}^{bc} 
= g^{db}(\boldsymbol{u})g^{ec}(\boldsymbol{u}) 
\partial_{a}\partial_{d}\partial_{e}G(\boldsymbol{u}) 
- \frac{\partial g_{ad}}{\partial\tilde{v}_{c}}(\boldsymbol{u})
g^{db}(\boldsymbol{u}) = 0, 
\end{equation}
for all $\boldsymbol{u}\in\mathcal{D}_{0}$, 
which implies that $\tilde{\boldsymbol{v}}$ is affine coordinates. 
In the derivation of (\ref{connection}), we have used the chain rule 
\begin{equation}
\frac{\partial g_{ad}}{\partial\tilde{v}_{c}}(\boldsymbol{u})  
= \frac{\partial g_{ad}}{\partial u^{e}}
\frac{\partial u^{e}}{\partial\tilde{v}_{c}}
= \partial_{a}\partial_{d}\partial_{e}G(\boldsymbol{u})
g^{ec}(\boldsymbol{u}). 
\end{equation}
Thus, (\ref{continuous_DE2_tmp}) reduces to the simple expression 
\begin{equation} \label{continuous_DE2_tilde} 
v_{a} 
= \tilde{v}_{a} 
+ \frac{M}{2}\sum_{\alpha=1}^{K}
\frac{\partial^{2}\tilde{v}_{a}}{\partial {x^{\alpha}}^{2}} + o(L^{2}), 
\end{equation} 
with (\ref{transform}). 
Similarly, (\ref{continuous_DE1}) is transformed into  
\begin{equation} \label{continuous_DE1_tilde} 
u^{a}(\boldsymbol{x},t+1) 
= \partial^{a}F(\boldsymbol{v})  
+ \frac{M}{2}\sum_{\alpha=1}^{K}
\frac{\partial^{2}\tilde{u}^{a}}{\partial {x^{\alpha}}^{2}}
+ o(L^{2}), 
\end{equation} 
with $\tilde{\boldsymbol{u}}=\{\tilde{u}^{a}:a=1,\ldots,N\}$. 

We next reduce (\ref{continuous_DE2_tilde}) and (\ref{continuous_DE1_tilde}) 
to the single equation~(\ref{PDE}) for $\tilde{\boldsymbol{v}}$. 
The derivation of (\ref{dual_PDE}) is performed in the same manner, and 
therefore omitted. 
Expanding the first term on the RHS 
of (\ref{continuous_DE1_tilde}) with (\ref{continuous_DE2_tilde}) yields  
\begin{equation} \label{gradient_F} 
\partial^{a}F(\boldsymbol{v}) 
= \partial^{a}F(\tilde{\boldsymbol{v}}) 
+ \frac{M}{2}f^{ab}(\tilde{\boldsymbol{v}})
\sum_{\alpha=1}^{K}\frac{\partial^{2}\tilde{v}_{b}}{\partial {x^{\alpha}}^{2}} 
+ o(L^{2}). 
\end{equation}
We use the chain rule to calculate the second derivative of 
(\ref{gradient_F}) with respect to $x^{\alpha}$ up to the order $O(1)$, 
\begin{IEEEeqnarray}{rl} 
&\frac{\partial^{2}}{\partial {x^{\alpha}}^{2}}\partial^{a}F(\boldsymbol{v}) 
\nonumber \\ 
=& \frac{\partial^{2}}{\partial {x^{\alpha}}^{2}}
\partial^{a}F(\tilde{\boldsymbol{v}}) + O(L^{-2}) \nonumber \\ 
=& f^{ab}(\tilde{\boldsymbol{v}})
\frac{\partial^{2}\tilde{v}_{b}}{\partial {x^{\alpha}}^{2}} 
+ \frac{\partial f^{ab}}{\partial v_{c}}(\tilde{\boldsymbol{v}}) 
\frac{\partial\tilde{v}_{b}}{\partial x^{\alpha}}  
\frac{\partial\tilde{v}_{c}}{\partial x^{\alpha}} + O(L^{-2}).
\label{derivative_F} 
\end{IEEEeqnarray}
Substituting (\ref{gradient_F}) and (\ref{derivative_F}) with 
$\tilde{u}^{a}=\partial^{a}F(\boldsymbol{v})$ into 
(\ref{continuous_DE1_tilde}) yields 
\begin{equation} 
u^{a}(\boldsymbol{x},t+1) - u^{a}(\boldsymbol{x},t) 
= - A^{a} + M\mathfrak{C}^{a}(\tilde{\boldsymbol{v}})
+ o(L^{2}), 
\end{equation}
with (\ref{coupling_operator}) and 
$A^{a}=u^{a} - \partial^{a}F(\tilde{\boldsymbol{v}})$.  

In order to prove (\ref{PDE}), we shall show that $A^{a}$ is a conservative 
field with the potential~(\ref{potential}). 
Let $\tilde{\Gamma}\subset\tilde{\mathcal{D}}_{0}$ denote a smooth curve 
connecting two points $\tilde{\boldsymbol{v}}_{0}$ and 
$\tilde{\boldsymbol{v}}$. When the inverse mapping of 
$\tilde{\boldsymbol{v}}=\nabla G(\boldsymbol{u})$ maps the curve 
$\tilde{\Gamma}$ to a curve $\Gamma\subset\mathcal{D}_{0}$,  
the line integral of $A^{a}$ along the curve $\tilde{\Gamma}$ yields 
\begin{IEEEeqnarray}{rl} 
\int_{\tilde{\Gamma}}A^{a}d\tilde{v}_{a} 
=&\int_{\Gamma}\left\{
 u^{a} - \partial^{a}F(\nabla G(\boldsymbol{u}))
\right\}g_{ab}(\boldsymbol{u})du^{b} \label{integral} \\ 
=& V(\boldsymbol{\Phi}(\tilde{\boldsymbol{v}})) 
+ \mathrm{Const}, \label{potential_tmp}  
\end{IEEEeqnarray}
where the potential~$V(\boldsymbol{u})$ is given by (\ref{potential}).  
Expression~(\ref{potential_tmp}) implies that $A^{a}$ is a conservative force. 
\end{IEEEproof} 
The derivation of (\ref{potential_tmp}) was also presented in \cite{Yedla122}. 
However, Yedla et al.\ did not explicitly explain how they found the 
integral~(\ref{integral}). Theorem~\ref{theorem1} implies that the potential 
emerges naturally when we use the appropriate parameterization 
in terms of differential geometry, i.e.\ the affine coordinate systems. 

\section{Potential Theory}
We replace the difference on the left-hand side (LHS) of (\ref{PDE}) by 
the differentiation, 
with (\ref{transform}), to obtain the partial differential equation (PDE) 
\begin{equation}
\frac{\partial\tilde{v}_{a}}{\partial t} 
= g_{ab}(\boldsymbol{\Phi}(\tilde{\boldsymbol{v}}))\left\{
 - \frac{\partial}{\partial\tilde{v}_{b}}
 V(\boldsymbol{\Phi}(\tilde{\boldsymbol{v}}))  
 + M\mathfrak{C}^{a}(\tilde{\boldsymbol{v}})
\right\}, \label{phenomenological_system} 
\end{equation}
with (\ref{coupling_operator}), 
where we have ignored the $o(L^{2})$ term. This replacement might be 
justified except for initial iterations, since the state for large $L$ 
changes quite slowly as $t$ increases. The boundary 
condition $\tilde{\boldsymbol{v}}(\boldsymbol{x},t)
=\tilde{\boldsymbol{v}}_{\mathrm{G}}
=\nabla G(\boldsymbol{u}_{\mathrm{G}})$ is imposed on the boundary 
$\partial\mathcal{C}_{K}$ of 
the $K$-dimensional hypercube $\mathcal{C}_{K}=[-1,1]^{K}\subset\mathbb{R}^{K}$, 
since the boundary condition 
$\boldsymbol{u}(\boldsymbol{l},t)=\boldsymbol{u}_{\mathrm{G}}$ has been 
imposed for all $\boldsymbol{l}\in\mathbb{Z}^{K}\backslash[-L+1:L-1]^{K}$. 
Note that the state $\tilde{\boldsymbol{v}}_{\mathrm{G}}$ corresponds to 
the good stable solution of the potential~(\ref{potential}), i.e.\ 
$V(\boldsymbol{\Phi}(\tilde{\boldsymbol{v}}_{\mathrm{G}}))
=V(\boldsymbol{u}_{\mathrm{G}})$. 

In order to investigate the convergence property of the PDE~(\ref{PDE}),  
we define the energy functional of a vector field  
$\tilde{\boldsymbol{v}}(\boldsymbol{x})$ on $\mathcal{C}_{K}$ 
\begin{equation} \label{functional} 
H(\boldsymbol{v}) 
= \int_{\mathcal{C}_{K}}\left\{ 
 V(\boldsymbol{\Phi}(\tilde{\boldsymbol{v}})) 
 + \frac{M}{2}\sum_{\alpha=1}^{K}
 \frac{\partial\tilde{v}_{a}}{\partial x^{\alpha}}
 \frac{\partial\tilde{v}_{b}}{\partial x^{\alpha}}
 f^{ab}(\tilde{\boldsymbol{v}})
\right\}d\boldsymbol{x}.    
\end{equation}
It is straightforward to find that (\ref{phenomenological_system}) 
is represented by 
\begin{equation}
\frac{\partial\tilde{v}_{a}}{\partial t} 
= - g_{ab}(\boldsymbol{\Phi}(\tilde{\boldsymbol{v}}))
\frac{\delta H}{\delta\tilde{v}_{b}}(\tilde{\boldsymbol{v}}), 
\end{equation}
where $\delta H/\delta\tilde{v}_{a}$ denotes the functional derivative 
of (\ref{functional}) for $\tilde{v}_{a}$. 
This expression indicates that the state 
$\tilde{\boldsymbol{v}}(\boldsymbol{x},t)$ moves in a direction where 
the energy functional~(\ref{functional}) decreases. In fact, 
a direct calculation implies that the energy functional~(\ref{functional}) 
is a Lyapunov functional for the dynamical 
system~(\ref{phenomenological_system}): 
\begin{equation}
\frac{dH}{dt}(\tilde{\boldsymbol{v}}) 
= - \int_{\mathcal{C}_{K}}
\frac{\delta H}{\delta\tilde{v}_{a}}(\tilde{\boldsymbol{v}})
\frac{\delta H}{\delta\tilde{v}_{b}}(\tilde{\boldsymbol{v}})
g_{ab}(\boldsymbol{\Phi}(\tilde{\boldsymbol{v}}))d\boldsymbol{x}\leq0, 
\end{equation}
where the equality holds only when $\delta H/\delta\tilde{v}_{a}=0$ is zero 
for all $a$. Here, we have used the non-negative definiteness of the 
Hesse matrix $\{g_{ab}(\boldsymbol{u})\}$. Furthermore, 
the energy functional~(\ref{functional}) is obviously bounded below. 
These observations imply that (\ref{functional}) is a Lyapunov functional.    
Thus, it is guaranteed that the state 
$\tilde{\boldsymbol{v}}(\boldsymbol{x},t)$ converges to a stationary solution  
$\tilde{\boldsymbol{v}}(\boldsymbol{x})$ as $t\to\infty$. 
Note that the convergence of 
$\tilde{\boldsymbol{v}}(\boldsymbol{x},t)$ implies the convergence 
$\lim_{t\to\infty}\tilde{\boldsymbol{u}}(\boldsymbol{x},t)=
\tilde{\boldsymbol{u}}(\boldsymbol{x})$. 
 
Instead of the stationary solution $\tilde{\boldsymbol{v}}(\boldsymbol{x})$ 
for (\ref{phenomenological_system}), 
we shall investigate properties of the stationary solution 
$\tilde{\boldsymbol{u}}(\boldsymbol{x})$ for (\ref{dual_PDE}), 
which is a solution to the boundary-value problem 
\begin{equation} \label{Newton_equation} 
M\tilde{\mathfrak{C}}_{a}(\tilde{\boldsymbol{u}})
= \frac{\partial}{\partial\tilde{u}^{a}}
\tilde{V}(\boldsymbol{\Psi}(\tilde{\boldsymbol{u}})), 
\end{equation}
with the boundary condition $\tilde{\boldsymbol{u}}(\boldsymbol{x})
=\boldsymbol{u}_{\mathrm{G}}$ for all 
$\boldsymbol{x}\in\partial\mathcal{C}_{K}$. 
In (\ref{Newton_equation}), we have ignored the $o(L^{2})$ term. 

In order to obtain an insight based on classical mechanics~\cite{Hassani12}, 
we assume $g_{ab}=\delta_{ab}$ and $K=1$. 
The differential equation~(\ref{Newton_equation}) with 
(\ref{dual_coupling_operator}) 
can be regarded as the Newton equation of motion: 
The state $\tilde{\boldsymbol{u}}(x^{1})$ is regarded as the position in 
$\mathcal{D}$ of a free particle with vanishing mass~$M$ at time~$x^{1}$, 
moving subject to the inverted potential 
$-\tilde{V}(\boldsymbol{\Psi}(\tilde{\boldsymbol{u}}))$. 
Note that $x^{1}$ is a temporal variable in this interpretation, whereas 
it has been introduced as the spatial variable for the SC system. 
The uniform solution 
$\tilde{\boldsymbol{u}}(x^{1})=\boldsymbol{u}_{\mathrm{G}}$ 
corresponds to the situation under which the particle continues to stay at  
the {\em unstable} solution $\boldsymbol{u}_{\mathrm{G}}$ of the {\em inverted} 
potential $-\tilde{V}(\boldsymbol{\Psi}(\tilde{\boldsymbol{u}}))$. 
On the other hand, non-uniform solutions to the situation under which 
the particle at the unstable solution $\boldsymbol{u}_{\mathrm{G}}$ 
at time~$x^{1}=-1$ moves somewhere, and comes back to 
$\boldsymbol{u}_{\mathrm{G}}$ at time~$x^{1}=1$. Vanishing mass $M$ implies that 
the velocity of the particle is infinitely quick. 
The conservation of energy implies that the latter situation never occurs if 
$\boldsymbol{u}_{\mathrm{G}}$ is the unique stable solution that minimizes  
the potential over all stationary solutions For any $g_{ab}$ and $K$,   
we follow this intuition to prove the following: 

\begin{theorem} \label{theorem2} 
If the boundary is fixed to the unique stable solution that 
minimizes the dual potential~(\ref{dual_potential}) over all stationary  
solutions, the uniform solution 
$\tilde{\boldsymbol{u}}(x)=\boldsymbol{u}_{\mathrm{G}}$ is the 
unique solution to the boundary-value problem~(\ref{Newton_equation}) with 
$\tilde{\boldsymbol{u}}(\boldsymbol{x})=\boldsymbol{u}_{\mathrm{G}}$ 
for all $\boldsymbol{x}\in\partial\mathcal{C}_{K}$ as $M\to0$. 
\end{theorem}

A result equivalent to Theorem~\ref{theorem2} was proved for a simple 
case~\cite{Ohashi13}. However, the methodology in this paper is more generic: 
Although the hypercube $\mathcal{C}_{K}$ is considered in this 
paper, the proof strategy for Theorem~\ref{theorem2} is applicable for any 
connected region with smooth boundaries. 
Theorem~\ref{theorem2} implies that, if $\boldsymbol{u}_{\mathrm{G}}$ satisfies 
the assumption in Theorem~\ref{theorem2},  
asymptotic performance of the BP 
algorithm for the GSC system converges to better performance 
$P(\boldsymbol{u}_{\mathrm{G}})$ for all positions, whereas it for the 
uncoupled system does to worse performance. 

\begin{IEEEproof}
Let us define a tensor $T_{\beta}^{\alpha}(\tilde{\boldsymbol{u}})$ as 
\begin{IEEEeqnarray}{rl}  
T_{\beta}^{\alpha}(\tilde{\boldsymbol{u}})  
=& M\delta^{\alpha\gamma}\frac{\partial\tilde{u}^{a}}{\partial x^{\gamma}} 
\frac{\partial\tilde{u}^{b}}{\partial x^{\beta}}
g_{ab}(\tilde{\boldsymbol{u}}) 
- \delta_{\beta}^{\alpha}\tilde{\mathfrak{L}}\left(
 \tilde{\boldsymbol{u}},
 \frac{\partial\tilde{\boldsymbol{u}}}{\partial\boldsymbol{x}} 
\right), \label{energy_tensor} 
\end{IEEEeqnarray}
where the dual Lagrangian~$\tilde{\mathfrak{L}}$ is given by 
\begin{equation} \label{dual_Lagrangian} 
\tilde{\mathfrak{L}}\left(
 \tilde{\boldsymbol{u}},
 \frac{\partial\tilde{\boldsymbol{u}}}{\partial\boldsymbol{x}} 
\right)
= \tilde{V}(\boldsymbol{\Psi}(\tilde{\boldsymbol{u}})) 
+ \frac{M}{2}\sum_{\alpha=1}^{K}
\frac{\partial\tilde{u}^{a}}{\partial x^{\alpha}}
\frac{\partial\tilde{u}^{b}}{\partial x^{\alpha}}
g_{ab}(\tilde{\boldsymbol{u}}).  
\end{equation}
Since (\ref{dual_Lagrangian}) is invariant under the 
translation of $\boldsymbol{x}$, we use Noether's theorem to find the 
conservation law 
\begin{equation} \label{general_conservation_law} 
T_{\beta}^{\alpha}(\tilde{\boldsymbol{u}}) 
= -A_{\beta}^{\alpha}(\boldsymbol{x}),  
\end{equation}
where $A_{\beta}^{\alpha}(\boldsymbol{x})$ is a divergence-free tensor 
independent of $\tilde{\boldsymbol{u}}$, i.e.\ 
$\partial A_{\beta}^{\alpha}/\partial x^{\alpha}=0$. It is straightforward to 
confirm the conservation law~(\ref{general_conservation_law}) 
by direct calculation. 

We shall prove the theorem by induction. First, let $K=1$. 
The conservation law~(\ref{general_conservation_law}) reduces to 
\begin{equation} \label{conservation_law} 
\frac{M}{2}\frac{d\tilde{u}^{a}}{dx^{1}}
\frac{d\tilde{u}^{b}}{dx^{1}}g_{ab}(\tilde{\boldsymbol{u}}) 
= \tilde{V}(\boldsymbol{\Psi}(\tilde{\boldsymbol{u}})) - A_{1}^{1}, 
\end{equation}
which corresponds to the conservation of energy. 
It is straightforward to find $A_{1}^{1}=\tilde{V}_{\mathrm{G}}
=\tilde{V}(\boldsymbol{\Psi}(\boldsymbol{u}_{\mathrm{G}}))$ as $M\to0$, 
by using the non-negative definiteness of $g_{ab}$, the boundary condition 
$\tilde{\boldsymbol{u}}(\pm1,t)=\boldsymbol{u}_{\mathrm{G}}$, and the fact that 
$d\tilde{\boldsymbol{u}}/dx$ must be infinite for all 
$\tilde{\boldsymbol{u}}\notin\mathcal{S}$ as $M\to0$.  

Let us prove $\tilde{\boldsymbol{u}}(0)=\boldsymbol{u}_{\mathrm{G}}$. 
Since the differential equation~(\ref{Newton_equation}) is invariant under 
the spatial reversal $\tilde{x}^{1}=-x^{1}$, the stationary solution has the 
symmetry $\tilde{\boldsymbol{u}}(-x^{1})=\tilde{\boldsymbol{u}}(x^{1})$. Thus, 
the differentiability of $\tilde{\boldsymbol{u}}$ for $x^{1}$ implies 
$d\tilde{\boldsymbol{u}}(0)/dx^{1}=\boldsymbol{0}$.  
Evaluating (\ref{conservation_law}) at the origin $x^{1}=0$, we find that 
$\tilde{\boldsymbol{u}}(0)=\boldsymbol{u}_{\mathrm{G}}$ must hold 
from the uniqueness of the stable solution $\boldsymbol{u}_{\mathrm{G}}$. 

The boundary-value problem on $[-1,1]$ 
has been decomposed into two small problems on $[-1,0]$ and $[0,1]$.  
From the spatial reversal symmetry of (\ref{Newton_equation}), the derivative 
of the stationary solution for $x^{1}$ is zero at the middle point of each 
interval. This observation and (\ref{conservation_law}) with 
$A_{1}^{1}=\tilde{V}_{\mathrm{G}}$ imply that $\tilde{\boldsymbol{u}}(x^{1})$ must 
be equal to $\boldsymbol{u}_{\mathrm{G}}$ at the middle points. Repeating this 
argument, we find that $\tilde{\boldsymbol{u}}(x^{1})=\boldsymbol{u}_{\mathrm{G}}$ 
must hold at countably infinite points. The continuity of the stationary 
solution implies that $\tilde{\boldsymbol{u}}(x^{1})=\boldsymbol{u}_{\mathrm{G}}$ 
for all $x^{1}\in[-1,1]$ is the unique solution. 

Next, assume $\tilde{\boldsymbol{u}}(\boldsymbol{x})=\boldsymbol{u}_{\mathrm{G}}$ 
for the $(K-1)$-dimensional case,  
and consider $K$-dimensional coupling. We focus on the 
$(K-1)$-dimensional hyperplane  
$\mathcal{P}_{1}=\{\boldsymbol{x}\in\mathcal{C}_{K}: x^{1}=0\}$ 
for the $x^{1}$-axis. Since the boundary-value problem~(\ref{Newton_equation}) 
is invariant under the reversal $\tilde{x}^{1}=-x^{1}$, we find the 
symmetry $\tilde{\boldsymbol{u}}(-x^{1},\ldots)
=\tilde{\boldsymbol{u}}(x^{1},\ldots)$, which implies 
$\partial\tilde{\boldsymbol{u}}/\partial x^{1}=\boldsymbol{0}$ 
on the hyperplane $\mathcal{P}_{1}$. Thus, the off-diagonal elements of 
$T_{\beta}^{\alpha}(\tilde{\boldsymbol{u}})$ is zero for $\alpha=1$ or $\beta=1$ 
on the hyperplane $\mathcal{P}_{1}$, so that $A_{\beta}^{\alpha}$ must also have 
the same structure on $\mathcal{P}_{1}$. This implies that 
the problem reduces to the $(K-1)$-dimensional case. By the assumption, 
we obtain $\tilde{\boldsymbol{u}}(\boldsymbol{x})=\boldsymbol{u}_{\mathrm{G}}$ 
on the hyperplane $\mathcal{P}_{1}$. 

Similarly, we find 
$\tilde{\boldsymbol{u}}(\boldsymbol{x})=\boldsymbol{u}_{\mathrm{G}}$ on the 
other hyperplanes $\mathcal{P}_{\alpha}=\{\boldsymbol{x}\in\mathcal{C}_{K}: 
x^{\alpha}=0\}$ for $\alpha=2,\ldots,K$. Thus, the boundary-value problem  
on the hypercube $\mathcal{C}_{K}$ has been decomposed into $2K$ small 
problems. Repeating the argument above implies that 
$\tilde{\boldsymbol{u}}(\boldsymbol{x})
=\boldsymbol{u}_{\mathrm{G}}$ must hold on the hypercube $\mathcal{C}_{K}$, 
since $\tilde{\boldsymbol{u}}(\boldsymbol{x})$ is continuous. 
\end{IEEEproof}

\section*{Acknowledgment}
The work of K.~Takeuchi was in part supported by the Grant-in-Aid for 
Young Scientists~(B) (No.~23760329) from JSPS, Japan.

\bibliographystyle{IEEEtran}
\bibliography{IEEEabrv,kt-itw2013}

\begin{thebibliography}{10}
\providecommand{\url}[1]{#1}
\csname url@samestyle\endcsname
\providecommand{\newblock}{\relax}
\providecommand{\bibinfo}[2]{#2}
\providecommand{\BIBentrySTDinterwordspacing}{\spaceskip=0pt\relax}
\providecommand{\BIBentryALTinterwordstretchfactor}{4}
\providecommand{\BIBentryALTinterwordspacing}{\spaceskip=\fontdimen2\font plus
\BIBentryALTinterwordstretchfactor\fontdimen3\font minus
  \fontdimen4\font\relax}
\providecommand{\BIBforeignlanguage}[2]{{%
\expandafter\ifx\csname l@#1\endcsname\relax
\typeout{** WARNING: IEEEtran.bst: No hyphenation pattern has been}%
\typeout{** loaded for the language `#1'. Using the pattern for}%
\typeout{** the default language instead.}%
\else
\language=\csname l@#1\endcsname
\fi
#2}}
\providecommand{\BIBdecl}{\relax}
\BIBdecl

\bibitem{Kudekar11}
S.~Kudekar, T.~Richardson, and R.~Urbanke, ``Threshold saturation via spatial
  coupling: Why convolutional {LDPC} ensembles perform so well over the
  {BEC},'' \emph{{IEEE} Trans. Inf. Theory}, vol.~57, no.~2, pp. 803--834, Feb.
  2011.

\bibitem{Kasai11}
K.~Kasai and K.~Sakaniwa, ``Spatially-coupled {MacKay-Neal} codes and
  {Hsu-Anastasopoulos} codes,'' \emph{IEICE Trans. Fundamentals}, vol. E94-A,
  no.~11, pp. 2161--2168, Nov. 2011.

\bibitem{Takeuchi11}
K.~Takeuchi, T.~Tanaka, and T.~Kawabata, ``Improvement of {BP}-based {CDMA}
  multiuser detection by spatial coupling,'' in \emph{Proc. 2011 IEEE Int.
  Symp. Inf. Theory}, Saint Petersburg, Russia, Aug. 2011, pp. 1489--1493.

\bibitem{Takeuchi122}
------, ``Performance improvement of iterative multiuser detection for large
  sparsely-spread {CDMA} systems by spatial coupling,'' \emph{{\rm submitted
  to} IEEE Trans. Inf. Theory}, 2012, [Online]. Available:
  http://arxiv.org/abs/1206.5919.

\bibitem{Schlegel11}
C.~Schlegel and D.~Truhachev, ``Multiple access demodulation in the lifted
  signal graph with spatial coupling,'' in \emph{Proc. 2011 IEEE Int. Symp.
  Inf. Theory}, Saint Petersburg, Russia, Aug. 2011, pp. 2989--2993.

\bibitem{Krzakala12}
F.~Krzakala, M.~M\'ezard, F.~Sausset, Y.~F. Sun, and L.~Zdeborov\'a,
  ``Statistical-physics-based reconstruction in compressed sensing,''
  \emph{Phys. Rev. X}, vol.~2, pp. 021\,005--1--18, May 2012.

\bibitem{Donoho12}
D.~L. Donoho, A.~Javanmard, and A.~Montanari, ``Information-theoretically
  optimal compressed sensing via spatial coupling and approximate message
  passing,'' in \emph{Proc. 2012 IEEE Int. Symp. Inf. Theory}, Boston, MA, USA,
  Jul. 2012, pp. 1231--1235.

\bibitem{Hassani12}
S.~H. Hassani, N.~Macris, and R.~Urbanke, ``Chains of mean field models,''
  \emph{J. Stat. Mech.}, no.~2, p. P02011, Feb. 2012.

\bibitem{Takeuchi121}
K.~Takeuchi, T.~Tanaka, and T.~Kawabata, ``A phenomenological study on
  threshold improvement via spatial coupling,'' \emph{IEICE Trans.
  Fundamentals}, vol. E95-A, no.~5, pp. 974--977, May 2012.

\bibitem{Yedla122}
A.~Yedla, Y.-Y. Jian, P.~S. Nguyen, and H.~D. Pfister, ``A simple proof of
  threshold saturation for coupled vector recursions,'' in \emph{Proc. IEEE
  Inf. Theory Workshop}, Lausanne, Switzerland, Sep. 2012.

\bibitem{Kudekar12}
S.~Kudekar, T.~Richardson, and R.~Urbanke, ``Wave-like solutions of general
  one-dimensional spatially coupled systems,'' \emph{\rm [Online]. Available:
  http://arxiv.org/abs/1208.5273.}, 2012.

\bibitem{Ohashi13}
R.~Ohashi, K.~Kasai, and K.~Takeuchi, ``Multi-dimensional spatially-coupled
  codes,'' \emph{{\rm accepted for publication in} Proc. 2013 IEEE Int. Symp.
  Inf. Theory}, 2013.

\end{thebibliography}

\end{document}